# Integrated optical devices based on broadband epsilon-near-zero meta-atoms

Lei Sun[1], Kin Wah Yu[2], and Xiaodong Yang[1*]

[1]*Department of Mechanical and Aerospace Engineering, Missouri University of Science and Technology*
*Rolla, MO 65409, USA*
[2]*Department of Physics, The Chinese University of Hong Kong*
*Shatin, N.T., Hong Kong*
[*]*Corresponding author: yangxia@mst.edu*

Compiled May 23, 2012

We verify the feasibility of the proposed theoretical strategy for designing the broadband near-zero permittivity (ENZ) metamaterial at optical frequency range with numerical simulations. In addition, the designed broadband ENZ stack are used as meta-atoms to build functional nanophotonic devices with extraordinary properties, including an ultranarrow electromagnetic energy tunneling channel and an ENZ concave focusing lens. © 2012 Optical Society of America

Recently, materials with near-zero permittivity (ENZ) have been widely investigated in theory and engineering due to their anomalous electromagnetic properties. Owing to the exotic features, ENZ materials possess a great deal of exciting applications, such as directive antenna and waveguide [1,2], electromagnetic transparency and cloaking design [3–6], radiation phase pattern converter [7,8], and electromagnetic energy squeezing and tunneling [9–11]. Moreover, the application about electromagnetic energy squeezing and tunneling has been experimental demonstrated in microwave frequency regime [12]. On the other hand, with the development of device nanofabrication, the design of optical ENZ materials has been recently proposed by several research groups [13, 14].

Previously, we have proposed a theoretical strategy to design an anisotropic broadband ENZ stack, where the effective permittivity can be near-zero in a wide frequency range, based on the Bergman spectral representation of effective permittivity [15]. The broadband ENZ stack is made of a multilayer structure, with each layer composed by metallic inclusions with the permittivity of $\varepsilon_1(\omega)$ embedded in a dielectric host with the permittivity of $\varepsilon_2$. Along the normal direction of the multilayer structure, the effective permittivity of the stack ($\varepsilon_e$) can be described as a function of a $s$-parameter

$$\varepsilon_e(s) = \left[\sum_{i=1}^{N} \frac{d_i}{\varepsilon_2(1 - f_i/s)}\right]^{-1}, \quad (1)$$

which is related to the thickness of each layer ($d_i$) and the metallic inclusion filling ratio ($f_i$) in each layer. The $s$-parameter is defined as $s = \varepsilon_2/(\varepsilon_2 - \varepsilon_1)$. On the other hand, according to the Bergman spectral representation, the effective permittivity of the stack can be mathematically characterized as

$$\varepsilon_e(s) = \varepsilon_2\left[1 - \sum_{i=1}^{N} \frac{F_i}{s - s_i}\right], \quad (2)$$

with a pair of spectral factors: the singularity ($s_i$) and the spectral density ($F_i$) corresponding to the operating frequency range and the desired effective permittivity. For the same ENZ stack, Eqs. (1) and (2) should be equal to each other. Therefore, the physical structure of the stack ($d_i$, $f_i$) can be retrieved from the spectral factors with an inverse algorithm [15]. The theoretical strategy has been demonstrated by several schematic examples in quasi-static condition. However, regarding practical applications, an examination in full-wave condition should be under consideration.

In this letter, we perform numerical simulations to verify the feasibility of our theoretical design strategy in full-wave condition with realistic materials, and demonstrate functional nanophotonic devices based on the designed broadband ENZ stack as the meta-atom. According to the theoretical strategy, we design a five-layer ENZ stack, depicted in the insert of Fig. 1, with operating frequency ranging from 439.3 THz to 472.2 THz. Silver is chosen as the metallic inclusion, which follows the Drude model as $\varepsilon_1(\omega) = \varepsilon_\infty - \omega_p^2/(\omega(\omega + \mathrm{i}\gamma))$, with $\varepsilon_\infty = 5.0$, $\omega_p = 1.38 \times 10^{16}$ rad/s, and $\gamma = 5.07 \times 10^{13}$ rad/s. While the dielectric host is silica with the permittivity of $\varepsilon_2 = 2.10$. The ENZ stack has the dimension of $x \times y \times z = 5 \times 5 \times 20\,\mathrm{nm}^3$. The thickness of each layer (normalized by the total thickness of the stack) and the metallic inclusion filling ratio in each layer are exactly calculated from the inverse algorithm and list in Table 1.

The effective permittivity of the ENZ stack is retrieved from its reflectance $S_{11}$ and transmittance $S_{21}$, according to the algorithm in Ref. [16]. The $S_{11}$ and $S_{21}$ are obtained from finite-difference time-domain (FDTD) nu-merical simulations. In the simulation, each layer is regarded as a homogenous and isotropic medium with an effective permittivity of

$$\varepsilon_e^{(i)} = f_i \varepsilon_\infty + (1-f_i)\varepsilon_2 - \frac{f_i \omega_p^2}{\omega(\omega + \mathrm{i}\gamma)}, \qquad (3)$$

based on the simple mixing rule. Figure 1 plots the real part and the imaginary part of the effective permittivity retrieved from the numerical simulation and the theoretical analysis based on Eq. (1). Clearly, the numerically retrieved real part of the effective permittivity varies around zero with small fluctuations (about ±0.05), covering the designed operating frequency range. According to Eq. (3), the properly designed metallic inclusion filling ratio in each layer ensures the zero effective permittivity around a specified operating frequency range. In this example (Table 1), the effective permittivity of each layer equals to zero at the frequency of 437.5 THz, 447.5 THz, 455.9 THz, 464.3 THz, and 474.3 THz respectively. The corresponding thickness of each layer then modifies the interactions between the adjacent layers, which finally results in a broadband ENZ response for the whole multilayer stack. Moreover, it is noticed that the numerically retrieved curve slightly increases with respect to the frequency, which is a little different from the theoretical curve. This difference is due to the phase variation of the reflected electromagnetic wave caused by individual layer at different frequency, which is not included in the theoretical design. However, the four peaks caused by the resonances at the interfaces between adjacent layers are predicted by the numerical simulation, which matches the theoretical design well. Furthermore, the dramatic changes of the effective permittivity from the inside to the outside of the operating frequency range are all clearly displayed in the numerically retrieved curve. Finally, it is noted that the reflection of the electromagnetic wave at the surface of the ENZ stack also leads to a higher optical loss retrieved from the numerical simulation.

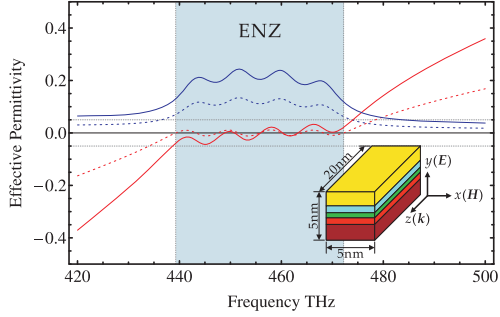

Fig. 1. (Color online) The real part and the imaginary part of the effective permittivity based on the FDTD simulation and the theoretical analysis from Eq. (1). The insert indicates the structure geometry of the broadband ENZ meta-atom stack including five layers. The theoretical effective permittivity is plotted as dashed curves, while the numerically retrieved permittivity is denoted as solid curves. In addition, the real part is shown in red color and the imaginary part is shown in blue color. The light blue color region represents the ENZ operating frequency range.

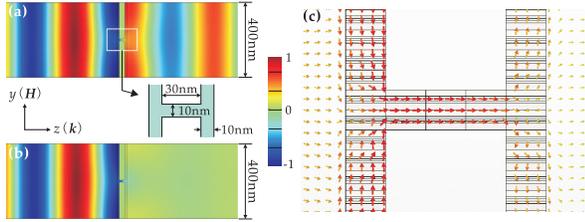

Fig. 2. (Color online) (a) The electromagnetic energy can be squeezed and tunneled through a H-shape channel made from the ENZ meta-atom stacks, but (b) cannot be transmitted through a silica channel with the same geometry, where the magnetic field distribution is displayed at 455 THz. (c) The energy flow distribution near the central channel region illustrates the squeezing and tunneling phenomenon.

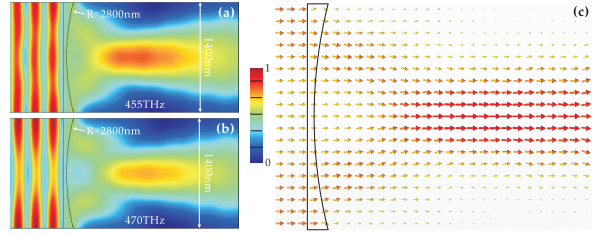

Fig. 3. (Color online) The electromagnetic waves can be focused by a concave lens made from the ENZ meta-atom stacks at (a) 455 THz and (b) 470 THz, where the magnetic field intensity distribution is displayed. (c) The energy flow with respect to the result of (a) clearly indicates the focusing of the electromagnetic energy. Due to the large quantity of the ENZ stacks within the concave lens, the detailed structure of the lens is not displayed here.

Table 1. Thickness and metallic inclusion filling ratio of each layer within the stack

| Layer | $d_i$ | $f_i$ |
|-------|----------|-----------|
| 1 | 0.314146 | 0.0942215 |
| 2 | 0.114413 | 0.0991492 |
| 3 | 0.108164 | 0.103449 |
| 4 | 0.11842 | 0.107877 |
| 5 | 0.344856 | 0.113305 |

The broadband ENZ stack can be considered as a deep subwavelength meta-atom to build functional nanophotonic devices with extraordinary optical properties, such as the ultranarrow electromagnetic energy squeezing and tunneling channel and the ENZ concave focusing lens. Figure 2 demonstrates an example about the electro-magnetic energy squeezing and tunneling through an ultranarrow channel constructed from broadband ENZ meta-atom stacks. Depicted in Fig. 2(a), a two dimensional H-shape channel is composed by the ENZ stacks and connected by silica waveguides on both sides. The arms of the channel are 400 nm in height and 10 nm in width, while the central tunnel is 10 nm in height and 30 nm in length. The space between the two arms of the channel is filled with optical absorbers in order to block the unwanted electromagnetic wave. The waveguides are sealed by perfect electric conductor (PEC) materials at the upper and lower boundaries to confine the electromagnetic energy. An electromagnetic wave at the frequency of 455 THz propagates in the waveguide from the left to the right, where the retrieved effective permittivity is $\varepsilon_e = -0.014 + 0.21i$ according to Fig. 1. Due to the impedance mismatch, one part of the electromagnetic wave is reflected back by the left surface of the channel and interferes with the incident wave. The rest of the incident electromagnetic wave is squeezed by the ENZ channel and transmitted through the central tunneling channel, then transferred to the waveguide at the right side. This phenomenon is caused by the effectively "infinite" phase velocity of the electromagnetic wave propagating inside the ENZ channel. Contrary to this exotic phenomenon, the same electromagnetic wave is totally reflected back by a silica H-shape channel with the same geometry, as displayed in Fig. 2(b). To illustrate a clear picture about the energy squeezing and tunneling, Fig. 2(c) plots the energy flow distribution near the central channel. It can be seen how the electromagnetic energy is coupled in from the left surface of the channel, squeezed into the central tunneling channel, and then reformed and released from the right side of the channel. It is noticed that due to the optical loss and the unavoidable reflection, the electromagnetic energy cannot be fully tunneled through the narrow central channel.

The ENZ material can also be used to focus the electromagnetic energy by manipulating the phase pattern of the electromagnetic wave. Since the designed ENZ stack possesses a broadband response, it can be used to focus the electromagnetic energy in a wide frequency range. In Fig. 3, a two dimensional concave focusing lens composed by the broadband ENZ meta-atom stacks is demonstrated, with the height of 1400 nm and the radius of curvature of 2800 nm. The concave lens is fixed in the end of a silica waveguide with the same height on the left, where the waveguide is sealed by PEC materials at the upper and lower boundaries. In Fig. 3(a), an electromagnetic wave at 455 THz in frequency couples to the concave lens from the waveguide and gets strongly focused, where the retrieved effective permittivity is $\varepsilon_e = -0.014 + 0.21i$ according to Fig. 1. The similar result can also be obtained at the frequency of 470 THz in Fig. 3(b), where the retrieved effective permittivity is $\varepsilon_e = 0.0034 + 0.17i$. Although the focusing is better at a higher frequency due to the shorter wavelength, the focal points are almost at the same location since there is almost no dispersion in ENZ materials. Figure 3(c) plots the energy flow distribution corresponding to the optical field of Fig. 3(a), where the electromagnetic energy focusing effect is illustrated clearly.

In conclusion, the feasibility of the theoretical strategy for design the broadband ENZ metamaterial is verified by full-wave numerical simulations. Moreover, the designed broadband ENZ meta-atom stack can be used to build functional nanophotonic devices in order to control the electromagnetic energy in exotic ways over a wide frequency range. It is worth noting that the current dimensions of the broadband ENZ meta-atom stack can be extended to larger sizes within the subwavelength limit, so that realistic experiments can be considered.

This work was partially supported by the Department of Mechanical and Aerospace Engineering, the Materials Research Center, the Intelligent Systems Center, and the Energy Research and Development Center at Missouri S&T, the University of Missouri Research Board, and the Ralph E. Powe Junior Faculty Enhancement Award. The authors acknowledge Y. He for his useful suggestions about this work.